\documentclass[12pt]{article}

\usepackage{scicite}

\usepackage{times}

\usepackage{graphicx}
\usepackage{dcolumn}
\usepackage{bm}
\usepackage{epsfig}
\usepackage{tcolorbox}
\usepackage{framed} 
\usepackage{tabularx}
\usepackage[normalem]{ulem}
\usepackage{amsmath}
\usepackage{amsfonts}
\usepackage{amssymb}
\usepackage{hyperref}

\newenvironment{sciabstract}{%
\begin{quote}{\bf Abstract.} \it}
{\end{quote}}

\title{Quantifying and predicting success in show business}

\author
{Oliver E. Williams,$^{1}$ Lucas Lacasa,$^{1}$ Vito Latora$^{1,2,3,4}$\\
\\
\normalsize{${}^{1}${\it School of Mathematical Sciences, Queen Mary University of London, London, E1 4NS, United Kingdom.}}\\
\normalsize{$^{2}${\it The Alan Turing Institute, The British Library, London NW1 2DB, UK,}}\\
\normalsize{${}^{3}${\it Dipartimento di Fisica ed Astronomia, Universit\`a di Catania and INFN,
I-95123 Catania, Italy.}}\\\normalsize{$^{4}${\it Complexity Science Hub Vienna (CSHV), Vienna, Austria.}}\\
\footnote{E-mails: o.e.williams@qmul.ac.uk | l.lacasa@qmul.ac.uk | v.latora@qmul.ac.uk}
}

\date{}

\begin{document} 


\baselineskip14pt


\maketitle


\begin{sciabstract}
Recent studies in the science of success have shown that the
highest-impact works of scientists or artists happen randomly and
uniformly over the individual's career. Yet in certain artistic
endeavours, such as acting in films and TV, having a job is perhaps
the most important achievement: success is simply making a living.  By
analysing a large online database of information related to films and
television we are able to study the success of those working in the
entertainment industry.  We first support our initial claim, finding
that two in three actors are ``one-hit wonders".  In addition we find
that, in agreement with previous works, activity is clustered in hot
streaks, and the percentage of careers where individuals are active is
unpredictable. However, we also discover that productivity in show
business has a range of distinctive features, which are
predictable. We unveil the presence of a rich-get-richer mechanism
underlying the assignment of jobs, with a Zipf law emerging for total
productivity.  We find that productivity tends to be highest at the
beginning of a career and that the location of the ``annus mirabilis''
-- the most productive year of an actor -- can indeed be predicted.
Based on these stylized signatures we then develop a machine learning
method which predicts, with up to 85\% accuracy, whether the annus
mirabilis of an actor has yet passed or if better days are still to
come.  Finally, our analysis is performed on both actors and actresses
separately, and we reveal measurable and statistically significant
differences between these two groups across different metrics, thereby
providing compelling evidence of gender bias in show business.
\end{sciabstract}

\newpage
\section{Introduction}

{\it ``It's feast or famine in showbiz." - Joan Rivers.} A sentiment
likely to be echoed by many would-be stars of the silver screen.  But
for those that feast the rewards are, at least thought to be, worth
the risk.  The so-called science of success has recently uncovered
many features of the careers of academics \cite{Sinatra}, artists
\cite{Galenson_02}, and all manner of other individuals whose output
can be effectively assessed over the course of their working life
\cite{Gilovich_1985,Rabin_2010,Brands_2006}.
In the world of scientific research it has revealed the 
unpredictability of the timing of an academics most impactful work
\cite{Sinatra}, showing that even such prestigious awards as Nobel
prizes are randomly distributed throughout the career of a scientist.
The anatomy of funding and collaborations in universities
has revealed ``rich clubs" of leading institutions, and suggested that
such patterns of collaborations 
contribute greatly to the success of these institutions, as 
measured in terms of over-attraction of available resources and
of breadth and depth of their research products \cite{Ma_2015}. 
Studies of innovation in industry across different countries have
found that the commercial success of manufacturing plants is far more
closely related to intra-group links than external ties
\cite{love_2001}. 
Strikingly, these features can be common across multiple areas; the
Matthew effect \cite{Merton_1968,Petersen_2011}, or {\it the rich get
  richer} phenomenon, and the recently discovered presence of ``hot
streaks'' \cite{Sinatra2}, are not restricted to isolated cases. With
regards to success, a great deal of work has been done in assessing
impact \cite{Sinatra,Hirsch_2005}, the distribution of standout or
landmark works \cite{simonton_1997,kozbelt_2008}, whether these are
related to the age of the individual in question
\cite{simonton_1988,lehman_1953}, how impact can be assessed in the
long term \cite{spitz_2014}, and even prediction of future successes
\cite{penner_2013,acuna_2012}. Indeed the fortunes of both films and
the actors and actresses that make them have been studied in some specific
ways \cite{pardoe_2008,mestyan_2013,gemser_2007,spitz_2014}. These
studies do not however address the question that interests those who
are not already on the higher rungs of the ladder of success: how can
one avoid the famine and build a sustainable career in acting?\\

\noindent The aim of this work is to use a data-driven approach in order to
define, quantify and even predict the success of actors and actresses
in terms of their ability to maintain a steady flow of jobs.  Drawing on
the International Movie Database (IMDb), an online database of
information related to films, television programs and home videos \cite{imdb}, we
have been able to study the careers of millions of actors from several
countries worldwide, from the birth of film in 1888 up to the present
day.  Each career is viewed as a profile sequence: the yearly time
series of acting jobs in films or TV series over the entire working
life of the actor or actress.  Note that all acting jobs are
considered, regardless of salary, role, screen time, or the impact of
the work.
The statistical analysis of such a large number of profile sequences
allows us to derive some general properties of the actors activity
patterns. In particular, we have looked at several quantities of
interest such as {\em career length}, {\em productivity} (defined as
the number of credit jobs in a year or in the entire career of an
actor) and position of the {\em annus mirabilis}, defined as the year
with the largest number of credited jobs. We have also explored
possible emergence of gender inequality in these properties.
\\

\noindent The first message that emerges from our quantitative
analysis is that one-hit wonders, i.e. actors whose career spans only
a single year, are the norm rather than the exception. Long career
lengths and high activity are found to be exponentially rare,
suggesting a scarcity of resources in the acting world.  We also see
that that this scarcity unequally applies to actors and actresses, providing
compelling evidence of gender bias.  Moreover, the total productivity
of an actor's career is found to be power-law distributed, with most
actors having very few jobs, while a few of them have more than a
hundred. This indicates a rich-get-richer mechanism underlying the
dynamics of job assignation, with already scarce resources being
allocated in a heterogeneous way.  All of this suggests that it is
activity that is truly important when measuring success in show
business. Only a select few will ever be awarded an Oscar, or have
their hands on the walk of fame, but this is not important to the
majority of actors and actresses who simply want to make a living.  It
is the continued ability to work, as opposed to prestige which is most
likely to ensure a stable career. For these reasons we propose that
predictions of success in show business should be focused on activity
and productivity.

\noindent Motivated by these results, we then address the questions
that interest the majority of working actors and actresses. Questions
such as ``am I going to get another paid job?'' or ``is this year
going to be my best?''.  We first show that efficiency, defined as the
ratio between the total number of active years and the career length,
is unpredictable, as there is no evident correlation between these two
things.  This is in line with recent studies \cite{Sinatra,Sinatra2}
pointing out that the most impactful pieces of work across artistic
and scientific disciplines usually happen randomly and uniformly over
an individual's career, and accordingly such achievements are
unpredictable.  Nevertheless, we here, surprisingly, find distinctive features 
in their temporal arrangement. In particular, we find that actor careers
are clustered in periods of high activity (hot streaks) combined with
periods of latency (cold streaks). 
Moreover, we discover that the most productive year (annus
mirabilis) for both actors and actresses is located towards the
beginning of their career, and that there are clear signals
preceding and following the location of the annus mirabilis of an
individual. Altogether, these unexpected results lead us to conclude that
prediction is possible in theory.  Finally, we validate this hypothesis
by building a statistical learning model which predicts the location
of the most productive year, finding that we can, with up to 85\%
accuracy, tell whether an actor's career has reached its most
productive year yet or not.

 \begin{figure}
\centering
\includegraphics[width=0.65\textwidth]{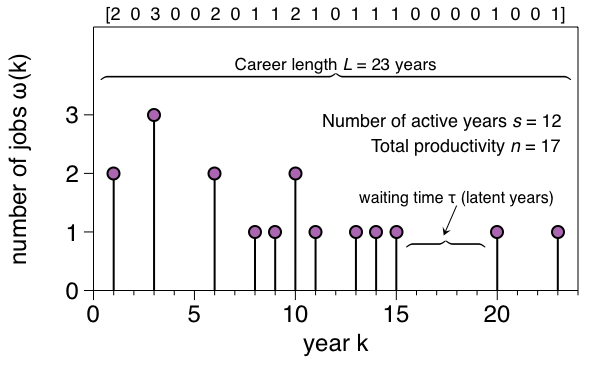}
\caption{{\bf Career activity pattern of an actor.} The yearly
  productivity of a given actor, measured as the total number of IMDb
  credited jobs in each year, is reported from the first to the last
  year of the actor activity. Shown is the case of an actor whose
  career spanned $L=23$ years and who was credited a cumulated $n=17$ different
  jobs in $s=12$ years.  From the yearly productivity we can construct
  the actor {\em profile sequence} $w_k$, with $k=1,\ldots,L$, shown in
  brackets above the plot, which can be modelled as a stochastic marked
  point process.  }
\label{fig:example}
\end{figure}

\section{Results}

We study the careers of $1,512,472$ actors and $896,029$
actresses as recorded on IMDb as of January 16th, 2016,
including careers stretching back to the first recorded movie in 1888.  The career of each
actor $a$ is characterised by his/her track record, which
consists of a set of pairs of numbers representing respectively each
 year when actor $a$ was credited in IMDb, and the number
of different credits in that year. As credits we count the
number of acting jobs in films and/or TV series. A sketch of the
typical activity pattern of an actor is reported in Figure
\ref{fig:example}, showing the yearly credits from the first to the last
year of thir career. Notice that there
are not only {\it active} years, where the actor has credited jobs in IMDb,
but also {\it latent} years with no recorded jobs. We therefore fill
the latent years with zeros and construct the {\it profile sequence}
$\{w_k\}_{k=1}^L$ of each actor $a$ as depicted in the top
part of Figure \ref{fig:example}. The quantity $w_k$ denotes the
actor's {\it local productivity} in year $k$, i.e. the number of credited
jobs in that year.
The {\it length} of an actor's career is defined as the number
of years between the first and the last active year (inclusive), and is denoted as $L$.
The total number of active years $s$ is from now on referred to
as the {\it activity} of an actor. Since a career can have latent years intertwined
with active ones we must have $s\leq L$, moreover $L-s$ is the number of
{\it latent years}. By definition we have: (i) $L \geq
1$, (ii) $s\geq 1$  and (iii) $s=1 \Leftrightarrow L=1$.
\\
Finally, we define the {\it total productivity} $n$ of an actor, as the
cumulated number of credited jobs, $n=\sum_{k=1}^L w_k$.
The {\it annus mirabilis} (AM) of
a given actor is defined as the year where the actor was credited with the largest number of
works in IMDb: $\text{AM}=m,$ where $m$ is such that $w_m=\max
\{w_k\}_{k=1}^L$. In the case that this $m$ is not unique we take the
final such year: $\text{AM} = \max \{m\}$.

\subsection{Career lengths and one-hit wonders}

\begin{figure*}[htp]
\centering
\includegraphics[width=0.32\textwidth]{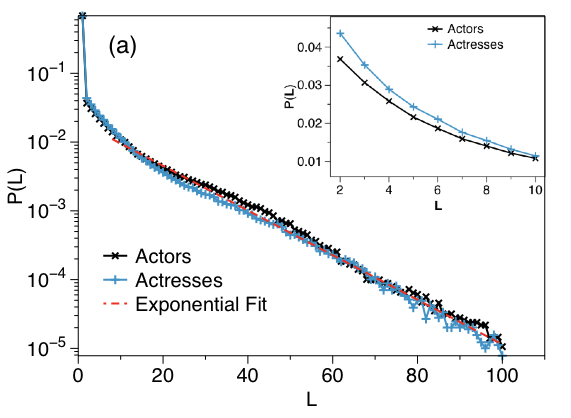}
\includegraphics[width=0.32\textwidth]{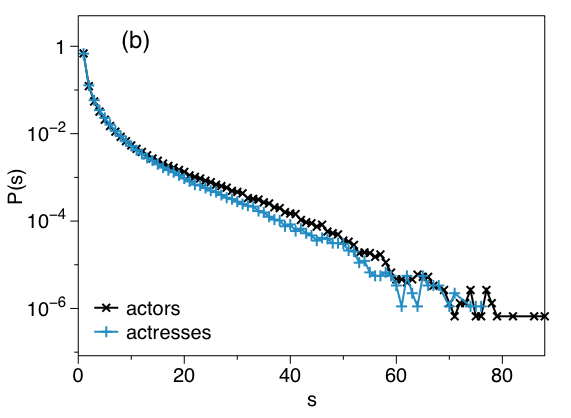}
\includegraphics[width=0.32\textwidth]{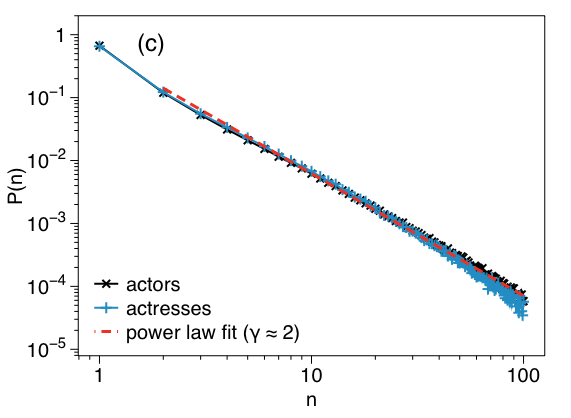}
\caption{{\bf Career length, activity and productivity distributions.}
{\bf (a)} The probability $P(L)$ that an actor or an actress has
  a career of length $L$, estimated by computing the frequency
  histogram of the number of years between the first and the last
  recorded entry on IMDb. $P(1)$ measures the abundance of ``one-hit
  wonders'', namely the actors or actresses with IMDB records in a
  single year. A zoom for $L\in[2,10]$ in the inset shows that careers
  extending between 2 and 10 years are proportionally more frequent in
  women than in men. {\bf (b)} Activity distribution $P(s)$ estimated
  by computing the frequency histogram of the number of working years
  within each career ($s\leq L$). Curves for actors and actresses are
  very similar and both exhibit a clear exponential tail, implying a `scarcity of resources'. {\bf (c)} Log-log plot of the total productivity distributions $P(n)$ for actors (black) and actresses (blue). Both curves decay as a power
  law $P(n)\sim n^{-\gamma}$, where $\gamma \approx 2$, revealing a
  Zipf law for the total number of acting jobs.}
\label{fig:length}
\end{figure*}

We start our analysis by exploring the statistics of the career length
$L$. In panel (a) of Figure \ref{fig:length} we plot in a
semi-log scale the empirical distribution of career lengths $P(L)$,
for both actors and actresses finding that the tail is well fitted by
an exponential distribution. By construction, $P(L=1)=P(s=1)$ and this
quantity represent the percentage of {\it one-hit wonders} i.e. of
actors whose career started and ended, according to IMDb, in
the same year. Interestingly, we find that the percentage of such
cases is extremely high (around $69\%$ for males and $68\%$ for
females) and deviates from the otherwise decaying exponential
distribution. This sharp deviation highlights that one-hit wonders are
not an exception in show business, but, on the contrary, are
the norm.
%
A zoom of the distribution in the range $L\in[2,10]$ is reported in the inset of
(a), revealing systematic differences between
actors and actresses, suggesting that it is consistently more
common to find (non-one-hit wonder) actresses with shorter
career lengths than actors.  We have indeed performed a model selection experiment which confirms that gender bias is statistically significant (see SI for details).\\

\noindent The empirical probability distribution of {\it activities}, displaying the
probability of sampling an actor that worked in $s$ years, is shown in
panel (b) of Figure \ref{fig:length} in a semi-log scale. Most of the actors and actresses are
only active in a single year ($s=1$), as by default $s=1\mapsto L=1$. 
The probability of finding actors with large activity, i.e. those that have
worked in many different years, decays exponentially
fast. This exponential decay mimics the similar decay in the
probability of finding long career lengths and altogether are the
basis for claiming a {\it scarcity of resources} in show business,
i.e. there are many more actors/actresses than job offers \cite{Mortensen_1986}. This lack of resources naturally leads to a question: how are they allocated? We address this question in the next section.

\subsection{Productivity and the rich gets richer phenomenon}

The right panel of figure \ref{fig:length} shows the empirical
distributions of total productivity $P(n)$, reporting the normalized
numbers of actors or actresses with $n$
 appearances in movies or TV series over
their careers. While the career length distribution $P(L)$ and the
activity distributions $P(s)$ are well
fitted in their tails by an exponential law, $P(n)$ decays
more slowly and can be fitted by a power law $P(n)\sim
n^{-\gamma}$ with exponent $\gamma\approx 2$. This emergence of a power
law shape in the distribution of total productivity 
implies that another scaling function emerges when the statistics are
computed in a rank-frequency plot. It is indeed well known \cite{lada}
that observing a power law distribution with exponent $\gamma$ for the
abundance of some variable is equivalent to obtaining a power law
scaling for the frequency of the variable that appears with rank $r$:
$f(r)\sim r^{-\alpha}$, where both scaling laws are mathematically
related via $\alpha=1/(\gamma-1)$. The celebrated Zipf law refers to
the particular case $\alpha\approx 1$, which is indeed the case
here. The fact that a Zipf law emerges for the rank-frequency
distribution of the total productivity of an actor may suggest a
mechanistic explanation for our observations.
Many different proposals for
the mechanism underpinning the emergence of a Zipf law, and
several names for the phenomenon itself, have been put forward
in various contexts, including the Simon-Yule process, the
mechanism of preferential attachment, the Matthew effect, the Gibrat
principle, rich-get-richer, etc.
\\ In this context, inspired by the
renowned preferential attachment paradigm as a generative model to
produce scale-free networks, we can easily explain the onset of a
power law distribution for the total productivity in terms of a
rich-get-richer heuristic: consider a generative model of an actor's
network, where nodes are actors and two actors are linked if they act
in the same film. Initially only a few actors are working, but as time
goes on new actors come into play. In this a way the new individuals work in
films together with other actors with a probability which is
proportional to the total productivity (i.e. the total number of
links) of the older actor-node. This preferential attachment
clearly expresses the rich-get-richer phenomenon by which actors that
work a lot will have a higher chance of working even more than actors
with low productivity. 
It is well known that such models generate
networks with power law degree distributions
\cite{BA,bollobas,vito,kleinberg}, i.e. power laws for the total
productivity. This result is not at all unexpected, after all, the
more well-known someone is, the more likely producers will
be to put him or her in their next film, if only for commercial
purposes. What is perhaps dramatic about this observation is that it
is well known that rich-get-richer effects are rather arbitrary and
unpredictable, as large hubs can evolve out of unpredictable and
random initial fluctuations which have been amplified, and not based
on any particular intrinsic fitness \cite{kleinberg} (such as acting
skills). Quoting Easly and Kleinberg: ``{\it if we could roll time
  back 15 years, and then run history forward again, would the Harry
  Potter books again sell hundreds of millions of copies, or would
  they languish in obscurity while some other works of children's
  fiction achieved major success?}''. As a matter of fact, 
  it seems likely that across different parallel universes popularity would 
still have a power law distribution, but it is far from
clear that the most popular actors would always be the
same. Interestingly, this hypothesis has recently been validated in an
online social experiment for the case of musical popularity
\cite{dodds}. In summary, productivity is probably the variable 
every actor aims to maximise, but these results suggest that 
boosting this is more of a network effect \cite{formula, new_science} 
than a consequence of `acting skills'.



\subsection{Efficiency is unpredictable} 
In figure \ref{fig:length} we observed that career length and activity
are variables which are both exponentially distributed, indicating a
scarcity of resources. In this section we further explore whether the
two quantities $L$ and $s$ are correlated. We first define 
an actor's {\it efficiency} as the ratio $s/L$ of active years
over the entire career, and we investigate how this 
efficiency is distributed. By construction, we have $s\leq L$ and
$L=1\to s=1$, and 
thus in this case $s/L=1$. As roughly $68\%$ of actors are
one-hit-wonders, we expect that the probability that an actor has
optimal efficiency $P(s/L=1)\approx P(L=1)$. However it is clear that
these are `pathological' cases as efficiency is not really well
defined for one-hit wonders.  In what follows, we therefore assume
that the case ($s=1$, $L=1$) is an outlier with regards to the
analysis of efficiency.
\\

\begin{figure}[htp]
\centering
\includegraphics[width=0.51\textwidth]{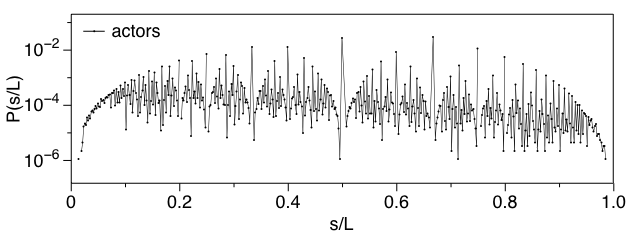}
\includegraphics[width=0.51\textwidth]{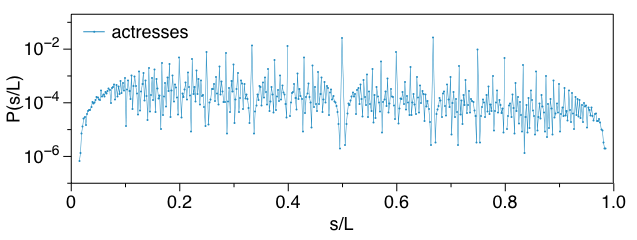}
\includegraphics[width=0.51\textwidth]{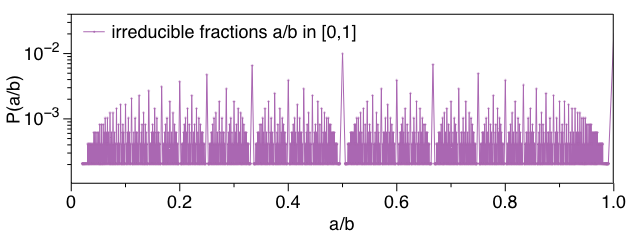}
\caption{{\bf Distribution of actor efficiency}.
  The probabilities of finding
  actors (top panel) and actresses (middle panel) with a given
  value of the ratio $s/L$ are shown as a function of $s/L$. 
  The curves have been computed after homogeneously binning the
  support [0,1] into 500 bins.
  The fractal shape of the curve is reminiscent of the binned density of
  irreducible fractions $a/b, \ b>a$, shown in the bottom panel.
  No obvious differences emerge here between actors and actresses.}
\label{fig:sL}
\end{figure}

\noindent In Figure \ref{fig:sL} we plot $P(s/L)$ on semi-log axes for actors
(top panel) and actresses (bottom panel). As might be expected, the
distribution decreases rapidly as $s/L$ approaches either zero or one,
suggesting that most actors and actresses have intermediate values of
efficiency (see SI for a discussion and a heuristic explanation of this phenomenon).
 The shape of $P(s/L)$ in the intermediate
range is fractal-like, which is due to the
fact that $s$ and $L$ are (small) integers and thus $s/L$ cannot take
arbitrary values in $[0,1]$. The fractal shape can actually be related to
the density of irreducible fractions over the integers, as depicted in
the bottom panel of Figure \ref{fig:sL}, and is not a property linked 
to the relation between $s$ and $L$. In other words, when this effect is
factored out, then $P(s/L)$ is essentially flat in the intermediate range.
Accordingly, correlations that emerge between the activity 
$s$ and the career length $L$ of an actor at intermediate ranges
seem to be related only to the fact that $s\leq L$. To further validate this
hypothesis, we performed a scatter plot of $s$ versus $L$ for all actors and
actresses, and computed the Pearson correlation coefficient.
This was then compared to the correlation coefficient of 
a null model generated by
randomly extracting values of $L$ and $s$ from the pool of
career profiles, ensuring that $L\geq s$.
For actors, we found that $s$ and $L$ correlate with a Pearson coefficient
$r\approx 0.67$, whereas in the null model we obtained 
$r_{\text{null}}\approx0.54$ (in the case of actresses $r\approx0.67$, to be compared with 
 $r_{\text{null}}\approx0.56$). As expected, $s$ and $L$ are
indeed correlated but almost all of those correlations can be
explained by a null model, concluding that for intermediate ranges
there are in practice not strongly influential additional correlations between length and activity: 
the activity of an actor cannot be predicted by their career length and
therefore we can conclude that the efficiency is an unpredictable
quantity.

\begin{figure}[htp]
\centering
\includegraphics[width=0.45\textwidth]{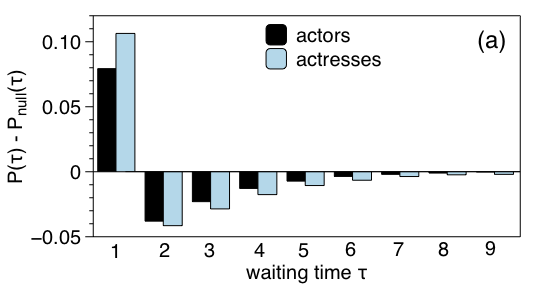}
\includegraphics[width=0.45\textwidth]{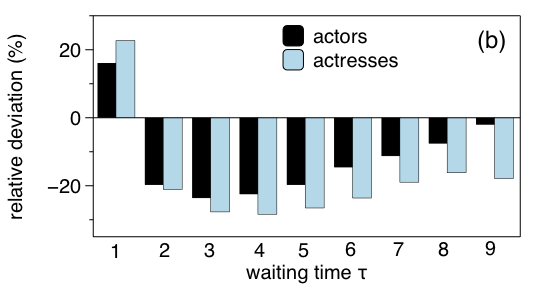}
\caption{{\bf Waiting time distribution}.
  {\bf (a)} Difference $P(\tau)-P_{\text{null}}(\tau)$ 
  between the waiting time distribution in the profile sequences and in
  a randomised null model, for actors (black bars) and
  actresses (blue bars). Systematically short waiting
  times, $\tau=1$, are overrepresented with respect to the null model,
  while the opposite is true for intermediate waiting times
  $\tau>1$. {\bf (b)} The percentage relative 
  difference $[P(\tau)-P_{\text{null}}(\tau)]\cdot100/P_{\text{null}}(\tau)$
  reveals a notable difference between actors and actresses.}
\label{fig:waiting}
\end{figure}

\subsection{Actors careers are clustered in hot and cold streaks} 
To understand the {\it temporal} arrangement of active years within
the profile sequence of a given actor, we now consider the statistics
of waiting times. A waiting time $\tau$ is defined as the time elapsed
(in years) between two active years (equivalently, a waiting time is a
collection of successive latent years), and its statistics provide a
classical way to analyse the presence of memory and bursts in time
series \cite{bak, corral}.  We have estimated the waiting time
distribution $P(\tau)$ for actors and actresses, 
discarding those with short career lengths, $L<10$ years, to
avoid a lack of statistics. To estimate this distribution, for each
actor (actress) we count how frequently one observes waiting times of
a certain duration $\tau$, and normalize the accumulated
frequencies. This process will inevitably introduce finite size
biases since, for short career lengths, we are more likely to find
short waiting times, simply because there is no room for long ones. For 
a proper comparison we therefore have also computed 
the distribution for a randomized
null model $P_{\text{null}}(\tau)$ where all of the profile sequences have
been shuffled (while keeping the first event $w_1$ and the last
event $w_L$ unaltered). A lack of temporal correlations would imply
$P_{\text{null}}(\tau)=P(\tau)$, whereas systematic differences
suggest the onset of temporal correlations in the activity of
actors. In panel (a) of Figure \ref{fig:waiting} we report the
difference $P(\tau)-P_{\text{null}}(\tau)$ as a function of $\tau$. 
\begin{figure}[htp]
\centering
\includegraphics[width=0.5\textwidth]{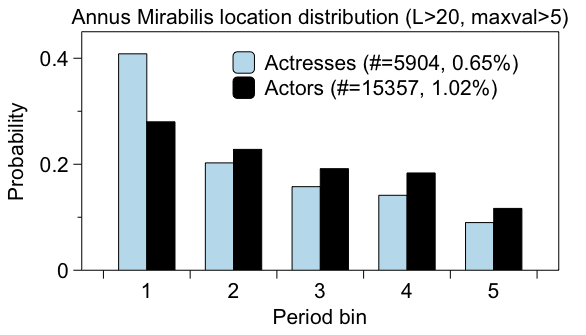}
\caption{{\bf Annus mirabilis tends to occur sooner rather than
    later.~~} Position of AM within an actor or actress's career,
  where the career length is binned into 5 bins in every case, to be
  able to compare profiles of different career lengths. We
  systematically find that the most probable location of the annus
  mirabilis is towards the beginning of a career.}
\label{fig:annus}
\end{figure}
For both actors and
actresses, we systematically find $P_{\text{null}}(\tau=1)<P(\tau=1)$,
and $P_{\text{null}}(\tau>1)>P(\tau>1)$, that is, active years are
more clustered than they would be by chance,
and hence the same is true of periods of inactivity.
This means that the profile sequence shows {\it
    clustering} and is composed of bursts of activity (hot streaks) where actors and actresses are more likely, than
  would be expected by chance, to work in a year if they worked
  the year before ($\tau=1$). This result is in agreement with recent findings in other creative jobs in
  science and art \cite{Sinatra2}.
  Additionally, these hot streaks are interspersed by
 abnormally long periods of latency (cold streaks) where authors are less likely than
  random to work in a given year if they did not work the year before
  ($\tau>1$).\\
Furthermore, to appropriately compare deviations from the null model 
for different waiting times, in panel (b) of Figure \ref{fig:waiting} 
 we plot the relative difference (in percentage)
$[P(\tau)-P_{\text{null}}(\tau)]\cdot100/P_{\text{null}}(\tau)$. We
find a substantial difference between actors and actresses: while
deviation from the null model decays for larger waiting times $\tau$
in the case of actors, for actresses this relative deviation is
maintained, pointing to a longer memory kernel, in turn suggesting that
having a period of latency is overall more detrimental for actresses
than for actors.

\subsection{Predicting the annus mirabilis}

It has recently been found that the most impactful publication 
that a scientist will produce is equally likely to occur at any stage of their 
career \cite{Sinatra}. Here we explore a related question in the
context of actors and actresses. Instead of impact, the 
indicator of success under study is productivity, as measured by the number of
credited works in IMDb. We concentrate on actors and actresses with
working lives extending beyond $L=20$ years, and define
the {\it annus mirabilis} (AM) of a given actor as the year  $k=y^*$ when the actor worked in the largest number of credited movies or TV series.
We restrict our reported results to those cases where there were at 
least 5 credited jobs in the AM, although other thresholds do
produce qualitatively similar results.
The subset of actors with $L>20$ and 
more than 5 acting jobs in the AM
 consists of 15357 actors (1.02\%) and 5904 actresses
(0.65\%). The large gender difference indicates that actors
tend to have more acting jobs than actresses.
\\
In Figure \ref{fig:annus} we plot the probability with which the AM will occur at each point within an actor or actress's career. 
To be able to compare these probabilities over careers of varying lengths,
 we have broken up each actor's time
series of $L$ years respectively into 5 bins (other segmentations produce qualitatively similar results).
 The plots consistently indicate that
the most probable location of the annus mirabilis is towards
the beginning of a career. Although the results are qualitatively
similar for male and
female actors, this bias is much more pronounced in the
case of actresses, further confirming the gender difference previously
observed.

\begin{figure*}[htp]
\centering
\includegraphics[width=0.48\textwidth]{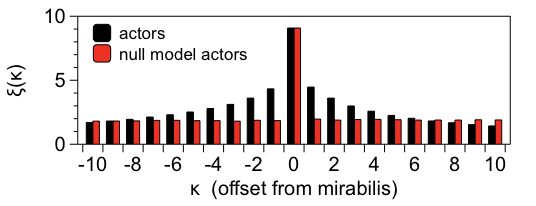}
\includegraphics[width=0.48\textwidth]{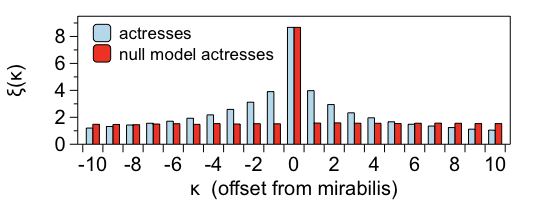}
\caption{{\bf The annus mirabilis is predictable.~~} The total number of
  acting jobs, $\xi(\kappa)$, averaged over all actors (left panel) and
  actresses (right panel), is reported as a function of the number of
  years $\kappa$ after or before the annus mirabilis. Only actors and actresses with
  a career lasting more than $L=20$ years and annus mirabilis with $w>5$
  acting jobs have been selected. In both cases, we observe a clear
  non-monotonic pattern, indicating that the annus mirabilis is
  either approaching or has just passed.  For comparison, we report in red the
  results obtained for a null model where the profile
  sequences of all actors and actresses have been shuffled. No pattern emerges in that case.}
\label{fig:early}
\end{figure*}

To study whether one can detect the imminent appearance of an
actor's annus mirabilis we have analysed, for both actors and
actresses, the average number of acting jobs before and after the
AM. In order to do this consistently, we initially perform a translation
$k \mapsto \kappa$
that aligns all profile sequences, so that the annus mirabilis $k=y^*$ all occur at
 $\kappa=0$. We then define:
$$\xi(\kappa)=\frac{1}{|{\cal A}|}\sum_{i=1}^{|{\cal A}|} w^{(i)}_{y^* + \kappa},$$
where $\kappa$ is the offset from the annus mirabilis and $|{\cal A}|$
is the size of the set of actors/actresses for which there exists a
profile sequence with an input at offset $\kappa$. In Figure
\ref{fig:early} we plot $\xi(\kappa)$,
showing that, on average, there
is a clear increase in the number of jobs preceding the AM and a clear
decrease immediately afterward. This pattern is absent in the corresponding 
null models obtained by shuffling the profile sequences (red bars). Such observed 
patterns can indeed be exploited for an early prediction of the
annus mirabilis.\\

Based on our observed distribution of jobs surrounding the annus mirabilis 
we propose a naive early-warning criterion: if the career sequence is 
non-monotonic around a value of $k$, i.e. if $w_k>w_{k-1}$ and $w_{k+1}<w_k$,
then the year $k$ is a good candidate for the annus mirabilis.
With this criterion in mind, one could ask the following question: given a
sample of an actor or actress's profile sequence, can we tell whether the
annus mirabilis has already passed or not?
\noindent Mathematically, the question above can be formalised as
follows: given a career sequence $(w_k)_{k=1}^L$ such that the maximal
total productivity occurs at time $k = y^*$,
consider a {\it truncated} sequence $\bar{w}_k=(w_k)_{k=1}^{T\leq
  L}$. We now wish to know if we can accurately asses whether $y^*
\in \{1,...,T\}$ using only $\bar{w}_k$. This forms a binary
classification problem, in which $\bar{w}_k \in \mathcal{C}_1$ if $y^*
\notin \{1,...,T\}$ and $\bar{w}_k \in \mathcal{C}_2$
otherwise. 
Our naive criterion, as illustrated above, readily
provides the heuristic: $\bar{w}_k \in \mathcal{C}_1$ if $\bar{w}_k$
is monotonic, and $\bar{w}_k \in \mathcal{C}_2$ if not. When this
method is tested on an appropriately generated set $\mathcal{W}$ of
truncated sequences (see SI for details) we find that it is
correct $\sim 69.2\%$ of the times for actors, and $\sim 75.0\%$ of the
times for actresses. This model now forms a benchmark against which we
will test a more refined approach. The idea is to relax
our classification method by introducing some parameters which allow
for deviation from the rigid heuristic, then train those
parameters on some subset $\mathcal{T} \subsetneq \mathcal{W}$, and
subsequently test the trained model on the test set $\mathcal{W}
\setminus \mathcal{T}$.  To do this let us first define the function
\begin{equation}
	D \left( \bar{w}_k \right) = - \sum_{y=1}^{T-1} \min \left( 0, \bar{w}_{y+1} - \bar{w}_{y} \right).
\end{equation}
At each year $k$ the contribution to $D$ from that year is zero
if the total productivity in the subsequent year is larger.
This means that for a monotonically increasing
sequence $\bar{w}_k$, $D \left( \bar{w}_k \right) = 0$.
If productivity decreases from year $k$ to $k+1$, then $D$ 
will increase by a corresponding amount.

$D \left( \bar{w}_k \right)$ effectively measures how far the sequence $\bar{w}_k$ is 
from being monotonically increasing, thus we can use it to relax our naive heuristic 
by defining some threshold $d$ such
that the decision rule $C \left( \bar{w}_k , d \right)$ becomes
\[   
C \left(  \bar{w}_k, d \right) = 
	\begin{cases}
	\mathcal{C}_1 &\quad\text{if } D \left( \bar{w}_k \right) < d\\
	\mathcal{C}_2 &\quad\text{if } D \left( \bar{w}_k \right) \geq d.    
	\end{cases}
\]
This new classifier is more flexible than the naive heuristic as we have introduced a parameter $d$ which can now be optimised (trained) as follows: if we denote $C^* \left(  \bar{w}_k \right)$ as the true class of the sequence $\bar{w}_k$, then the optimal value of the parameter $d^*$ is the value of $d$ that minimises the following loss function
\begin{equation}
	L\left(\mathcal{T}, d \right) = - \sum_{\mathcal{T}} \delta \left( C \left(  \bar{w}_k , d \right), C^* \left(  \bar{w}_k \right) \right).
\end{equation}
Where $\delta(X,Y)$ yields one if $X = Y$ and 0 otherwise.
This value for $d^*$ is then used to classify the remaining 
sequences in $\mathcal{W} \setminus \mathcal{T}$.
The results of this testing on both actors and actresses
can be partially summarised by the two confusion matrices $\text{CO}_m$ (for actors) and $\text{CO}_f$ (for actresses):
\[
\text{CO}_m = 
\begin{bmatrix}
	33775 & 5659 \\
	10771 & 52000
\end{bmatrix}
,~
\text{CO}_f = 
\begin{bmatrix}
	12549 & 2593 \\
	3596 & 26682
\end{bmatrix}
\]
The classical metrics used to assess the performance of the
classifier, namely accuracy, precision, recall and the F1 score, are
summarised in Table \ref{table:confusion}. We find that the accuracies
of the prediction are $84\%$ and $86\%$ respectively, i.e. $\sim 10\%$
higher than those obtained using a naive heuristic.

\begin{table}
\centering
\begin{tabular}{|c|c|c|}
\hline
{\bf Quantity}& {\bf Actors}&{\bf Actresses}\\
\hline
Total $\mathcal{C}_1$& 44652 & 16145\\
Total $\mathcal{C}_2$& 57553 & 29275\\
\hline
Accuracy & 0.8405 & 0.8637\\
\hline
Precision & 0.8608 & 0.8287\\
\hline
Recall & 0.7575 & 0.7773\\
\hline
F1 score & 0.8058 & 0.8021\\
\hline
\end{tabular}
\caption{ \label{table:confusion} Performance metrics (accuracy,
  precision, recall and F1 score) of the proposed classification method
  for the prediction of the annus mirabilis.}
\end{table}

\section{Discussion}

In this work we have made use of the vast quantity of 
data presented by IMDb to explore, analyse and predict 
success on the silver screen. By studying the careers of 
 $1,512,472$ actors and $896,029$ actresses from 1888 
up to 2016, we have uncovered a number of distinctive
patterns that characterize various aspects of the film and
TV industries. Such patterns not only allow us to identify
qualities of individual actors or actresses working lives, but also
to gain a deeper insight 
into the mechanisms by which jobs are themselves assigned.
Based on our findings, we have then constructed a statistical 
learning model that predicts with high accuracy whether an actor or
actress is likely to have a bright future, or if the best days are,
unfortunately, behind them.

\noindent The analysis performed in the first part of our work
supports the following eight key observations: {\em (i)} One-hit wonders 
are the norm, rather than the exception.
This implies that productivity  is probably a variable more
closely related to success, rather than 
the importance or impact of particular jobs.
{\em (ii)} Resources are scarce: long careers and higher activities are 
exponentially rare, implying that there are far fewer jobs available 
than there are applicants.
{\em (iii)} A rich-get-richer mechanism underlies the way jobs are
assigned, as evidenced by the emergence of Zipf's law in the
distribution of both actors and actresses total number of credited
works. This fact suggests \cite{dodds} that high
productivity is likely to be a network effect \cite{formula,
  new_science} rather than purely based on merit, i.e. productivity
does not necessarily correlate with acting skills.
{\em (iv)} The efficiency of an actor (the ratio between active years and
length career) is unpredictable, a finding reminiscent of the
randomness discovered in \cite{Sinatra}.
{\em (v)} Careers are clustered in periods of high activity or
``hot-streaks'' (again in agreement with recent findings
\cite{Sinatra2}) interspersed with periods of latency (cold streaks),
an effect which is more pronounced for actresses.
%
%
{\em (vi)} The most productive years for both actors and actresses 
tend to be towards the start of their careers though, again, this
is more evident for actresses. Again, we see here evidence for 
gender bias: older actresses are far less likely to maintain their 
status in the acting world than men.
%
{\em (vii)} There are clear signals both preceding and following 
the most productive years of an actor or actresses career. 
{\em (viii)} Including the aforementioned points, we have found statistically 
significant differences between actors and actresses across a wide 
range of metrics, providing credible and convincing evidence of 
gender bias.

\noindent The second part of our work deals with the prediction of an actors
productivity. By utilising the observed patterns present across the
careers of both actors and actresses we have produced a statistical
learning model which, given a sample of a career, is able to predict
with roughly 85\% accuracy whether that career has peaked, or whether
there are better things in store for our budding star. \\


To conclude, a life in show business appears to be very different to
those of artists or scientists. Where recent works have found an
inherent unpredictability and randomness in the careers of academics
and creatives we have found predictability and patterns. It is now
natural to ask why these differences occur; why were we able predict
the fortunes of individuals in one area while it has been shown to be
impossible to do so elsewhere? 
Another natural and intriguing problem to investigate might be the 
following: suppose we have predicted that a certain actors 
annus mirabilis has passed. What then can the actor do
  to change their fortunes and bring about even greater success?
We hope that our methodology and the
results that we have obtained will contribute to the ongoing debate
surrounding the science of success \cite{formula}.  Given the scope of
our findings across the industry, we also wish that our work will be
of interest to those working within it.





\section*{Acknowledgments}
LL acknowledges support from EPSRC grant EP/P01660X/1.
  VL acknowledges support from EPSRC Grant EP/N013492/1.

\section*{Author contributions}
LL, VL and OW designed the study. OW and LL performed the data analysis. All authors interpreted results and wrote the paper.

\section*{Author declaration}
The authors declare no conflicts of interest. 

\section*{Supplementary materials}
Supplementary Text\\
Reference \textit{34}


\end{document}